\documentclass[12pt,a4paper]{article}

\usepackage{ajr}

\usepackage{defns}

\begin{document}

\title{Use the information dimension, not the Hausdorff}
\author{A.~J. Roberts\thanks{Dept.~Mathematics \& Computing, University
of Southern Queensland, Toowoomba, Queensland 4350, \textsc{Australia}. 
\protect\url{mailto:aroberts@usq.edu.au}}}

\date{Original version 1997, revised \today}

\maketitle

\begin{abstract}
Multi-fractal patterns occur widely in nature.  In developing new 
algorithms to determine multi-fractal spectra of experimental data I 
am lead to the conclusion that generalised dimensions~$D_q$ of order 
$q\leq0$, including the Hausdorff dimension, are effectively 
\emph{irrelevant}.  The reason is that these dimensions are 
extraordinarily sensitive to regions of low density in the 
multi-fractal data.  Instead, one should concentrate attention on 
generalised dimensions~$D_q$ for $q\geq 1$, and of these the 
information dimension~$D_1$ seems the most robustly estimated from a 
finite amount of data.
\end{abstract}

\tableofcontents

\section{Introduction}

The characterisation of spatial distributions in terms of fractal
concepts~\cite{Mandelbrot79, Feder88} is becoming increasingly
important.  In particular, many distributions in nature are found to
have the characteristics of a multi-fractal~\cite{Hentschel83,
Halsey86, Paladin87}: among many examples are galaxy
clustering~\cite{Borgani93, Martinez91}, strange
attractors~\cite{Procaccia88a}, fluid turbulence~\cite{Sreenivasan91},
percolation~\cite{Isichenko92}, the shapes of
neurons~\cite{Jelinek01, Jelinek04},
and plant distributions~\cite{Emmerson95} and shapes~\cite{Jones96}.

In application, methods for estimating fractal dimensions are often
unreliable.  One source of error lies in largely unknown biases
introduced by the finite size of data sets, addressed by
Grassberger~\cite{Grassberger88b}, and in the associated finite range
of length-scales inherent in gathered data.  In situations where
thousands or tens of thousands of data points are known such biases may
be minor; however, in some interesting problems, for example in the
spatial clustering of underwater plants~\cite{Emmerson95}, only of the
order of 100 data points are known and confidence in the fractal
characterisation may be misplaced.  We need to know more about factors
that cause errors in dimension estimates.

Section~\ref{ss2} discusses the sensitivity of the multiplicative
multi-fractal process to regions of very low probability (measure).
Since such regions only rarely contribute a data point, an experimental
sample cannot discern them but such regions do affect the generalised
dimensions.  Hence I argue that the determination from experimental
data of generalised dimensions,~$D_q$, for non-positive~$q$ is
meaningless; for $0<q<1$ computations are very sensitive to the sample;
and thus the most robust fractal dimension is the information
dimension~$D_1$.  The argument is supported in Section~\ref{ss3} by a
maximum likelihood method~\cite{Roberts95b} of estimating the
multi-fractal properties of a data set.  The method shows the enormous
sensitivity of~$D_q$ for negative~$q$.  In contrast the information
dimension is reliably estimated.

\section{Poor conditioning of generalised dimensions of negative order}
\label{ss2}

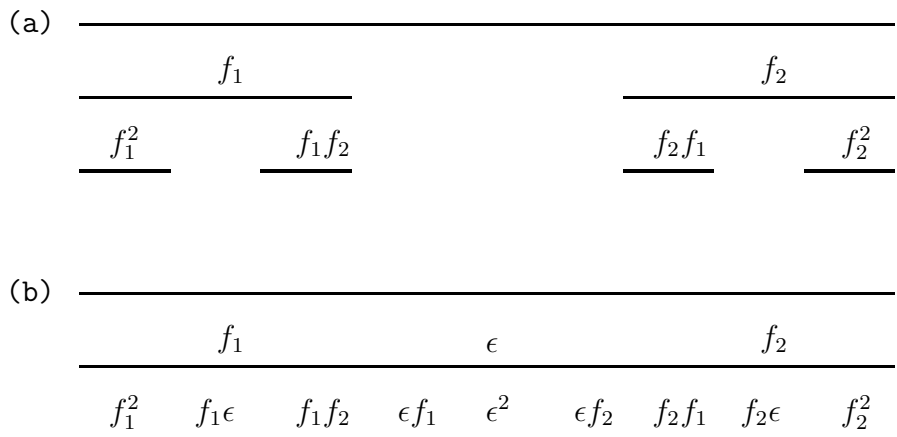
\begin{figure}[tbp]
	\centerline{{\tt    \setlength{\unitlength}{0.075em}
\begin{picture}(402,196)
\thinlines    \put(242,18){$\epsilon f_2$}
              \put(170,18){$\epsilon f_1$}
              \put(206,18){$\epsilon^2$}
              \put(310,18){$f_2\epsilon$}
              \put(87,18){$f_1\epsilon$}
              \put(206,46){$\epsilon$}
              \put(336,10){\line(-1,0){37}}
              \put(151,10){\line(1,0){111}}
              \put(77,10){\line(1,0){37}}
              \put(151,40){\line(1,0){111}}
              \put(40,10){\begin{picture}(333,62)
\thicklines       \put(311,8){$f_2^2$}
                  \put(234,8){$f_2f_1$}
                  \put(88,8){$f_1f_2$}
                  \put(12,8){$f_1^2$}
                  \put(278,38){$f_2$}
                  \put(56,38){$f_1$}
                  \put(333,0){\line(-1,0){37}}
                  \put(222,0){\line(1,0){37}}
                  \put(111,0){\line(-1,0){37}}
                  \put(0,0){\line(1,0){37}}
                  \put(222,30){\line(1,0){111}}
                  \put(0,30){\line(1,0){111}}
                  \put(0,60){\line(1,0){333}}
                  \end{picture}}
                  \put(40,120){\begin{picture}(333,62)
\thicklines       \put(311,8){$f_2^2$}
                  \put(234,8){$f_2f_1$}
                  \put(88,8){$f_1f_2$}
                  \put(12,8){$f_1^2$}
                  \put(278,38){$f_2$}
                  \put(56,38){$f_1$}
                  \put(333,0){\line(-1,0){37}}
                  \put(222,0){\line(1,0){37}}
                  \put(111,0){\line(-1,0){37}}
                  \put(0,0){\line(1,0){37}}
                  \put(222,30){\line(1,0){111}}
                  \put(0,30){\line(1,0){111}}
                  \put(0,60){\line(1,0){333}}
                  \end{picture}}
\thicklines   \put(10,67){(b)}
              \put(10,177){(a)}
\end{picture}}}
	\caption{schematic diagram of the first few stages in the
	multiplicative multi-fractal process to illustrate the sensitivity
	of the Hausdorff dimension~$D_0$ with respect to low density
	regions,~(b), as a perturbation of the same process with zero
	density regions,~(a).}
	\protect\label{ftern}
\end{figure}

For example, consider the Hausdorff dimension,~$D_0$, of multifractals 
generated by two different ternary multiplicative process.
\begin{itemize}
	\item Consider first the process shown in Figure~\ref{ftern}(a) 
	where an interval is divided into three thirds and the ``mass'' of 
	the original interval is assigned as follows: a fraction $f_1>0$ 
	to the left third; a fraction $f_2=1-f_1>0$ to the right third; 
	and none to the middle third.  Repeat this subdivision 
	recursively.  This generates a multiplicative multifractal whose 
	Hausdorff dimension of $D_0=\log_32=0.6309$ is precisely the same 
	as the Cantor set because there is no ``mass'' in the middle 
	thirds.

	\item Conversely, and perversely, consider the process shown in 
	Figure~\ref{ftern}(b) where for some small~$\epsilon$ the ``mass'' 
	is assigned as follows: a fraction $f_1>0$ is assigned to the left third; 
	a fraction $f_2>0$ is assigned to the rightmost third; and a small 
	fraction $\epsilon>0$ is assigned to the middle third (such that 
	$f_1+f_2+\epsilon=1$).  Repeat recursively.  This generates a 
	multiplicative multi-fractal whose Hausdorff dimension is $D_0=1$ 
	because there is ``mass'' everywhere along the whole interval!  
	Although the vast bulk of the ``mass'' can be covered by~$2^n$ 
	intervals of length~$3^{-n}$, we definitely do need~$3^n$ 
	intervals in order to ensure coverage of the thinly spread 
	``mass'' that fills most of the original interval.
\end{itemize}
The importance of this for the analysis of an experimental data set of
$N$~sampled points is that one cannot tell the difference from the data
between these two multi-fractal generating processes for an
$\epsilon=\ord{1/N}$.  Thus one cannot estimate the Hausdorff
dimension~$D_0$ with any accuracy since either answer, $0.6309$~or~$1$
could be correct.

\begin{figure}[tbp]
	\centerline{\includegraphics[width=0.95\textwidth]{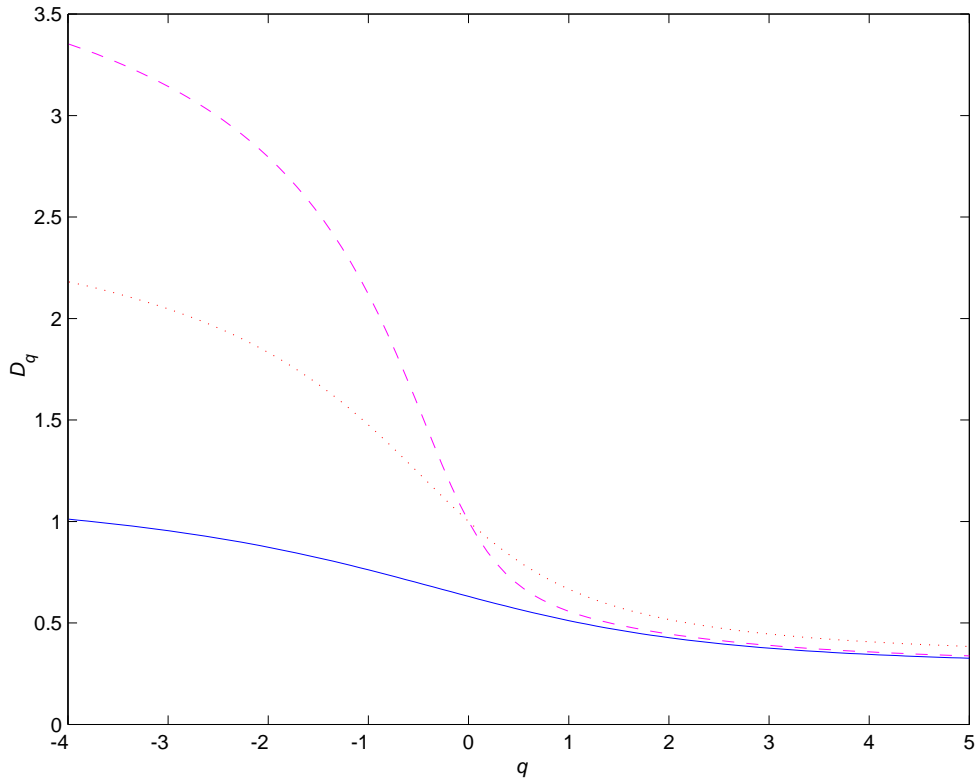}}
	\caption{multi-fractal generalised dimensions~$D_q$ for the 
	ternary multi-fractal process with $f_1=(1-\epsilon)/4$, 
	$f_2=(1-\epsilon)3/4$ and $\epsilon=0$ (solid), $0.01$ (dashed) 
	and $0.05$ (dotted).  This figure shows that $D_q$~for negative 
	order~$q$ is extraordinarily sensitive to small influences: the 
	curve of smaller~$\epsilon$ is the most changed.}
	\protect\label{fthe}
\end{figure}

Similar reasoning applies to generalised dimensions with negative~$q$.
Elementary arguments give that the generalised
dimensions~\cite{Hentschel83} of the multi-fractal generated by the
second process above are
\begin{equation}
	D_q=\left\{
	\begin{array}{ll}
		\frac{-1}{q-1}\log_3\left[f_1^q+f_2^q+\epsilon^q\right] & \mbox{if 
		}q\neq 1\,,  \\
		-\left[f_1\log_3f_1+f_2\log_3f_2+\epsilon\log_3\epsilon\right] & 
		\mbox{if }q=1\,.
	\end{array}\right.
\end{equation}
It is readily appreciated that for negative order~$q$ and
small~$\epsilon$, the term~$\epsilon^q$ inside the logarithm 
dominates the evaluation of the generalised dimension~$D_q$.  Hence, all
generalised dimensions for negative~$q$ are also extremely sensitive to
small~$\epsilon$.  In a data set obtained from experiments, one cannot
expect to distinguish between zero~$\epsilon$ and small non-zero
$\epsilon=\ord{1/N}$, and yet the generalised exponents and
multi-fractal spectrum are markedly different.  See Figure~\ref{fthe}
which plots the generalised dimensions for $f_1\approx1/4$,
$f_2\approx3/4$ and various small~$\epsilon$.

We can be more precise about the sensitivity to low density regions 
by computing the derivative of~$D_q$ with respect to~$\epsilon$.  For 
definiteness, suppose $f_1=\phi_1(1-\epsilon)$ and 
$f_2=\phi_2(1-\epsilon)$.  Then 
\begin{equation}
	\frac{\partial D_q}{\partial \epsilon}=\frac{-q}{q-1} 
	\frac{\epsilon^{q-1}-\left(\phi_1^q+\phi_2^q\right)(1-\epsilon)^{q-1}}%
{\log3\,\left[\epsilon^q+\left(\phi_1^q+\phi_2^q\right)(1-\epsilon)^q\right]}
\,.
	\label{ede}
\end{equation}
For small, but non-zero, $\epsilon\to 0$ this asymptotes to
\begin{equation}
	\frac{\partial D_q}{\partial\epsilon}\sim\frac{1}{\log3}\left\{
	\begin{array}{ll}
		\frac{q}{q-1} & \mbox{if }1<q\,,  \\
		\frac{q}{(1-q)(\phi_1^q+\phi_2^q)}\epsilon^{q-1} & \mbox{if 
		}0<q<1\,,  \\
		\frac{q}{1-q}\epsilon^{-1} & \mbox{if }\phantom{0<}q<0\,.
	\end{array}
	\right.
	\label{easy}
\end{equation}
This derivative is unbounded as $\epsilon\to0$ for $q<1$, and so any 
computation of~$D_q$ is only robust if $q\geq 1$.

The reason for this aberrant behaviour is clear.  With a finite number
of data points, it is impossible to tell the difference between truly
empty space and space which is visited so rarely that no data point
happens to fall within it.  That is, one cannot tell the difference
between empty space and space that should be filled in with very low
probability.  These differences dramatically affect the generalised
dimensions~$D_q$ for $q<1$.  Thus for any experimental data set:
\begin{itemize}
	\item  estimating~$D_q$ for $q\leq0$ is nonsense (including the Hausdorff 
	dimension);
	
	\item  estimates of~$D_q$ for small positive~$q$ are sensitive; and
	
	\item I only recommend the reporting of dimensions~$D_q$ 
	for $q\geq1$ as being robust.
\end{itemize}
Out of all the generalised dimensions for order $q\geq 1$, $D_1$~is 
most representative of the fractal as a whole.  For large order~$q$, 
the computation of~$D_q$ is determined only by the very ``densest'' 
regions of the multi-fractal and so is not representative of the whole 
fractal.  In the above multiplicative process,
\begin{displaymath}
	D_q\sim-\log_3\mbox{max}(f_1,f_2,\epsilon)
	\quad\mbox{as}\quad q\to\infty\,,
\end{displaymath}
showing that the large~$q$ behaviour is dictated by the one parameter 
of the process that determines the character of the very densest 
clusters in the fractal.  The very dense clusters occur rarely in the 
fractal; they have low fractal dimension as seen in the low~$f$ value 
typically associated with low values of~$\alpha$ in the multi-fractal 
spectrum.  Because of this rareness, the computation from experimental 
data of~$D_q$ for large positive order~$q$ is unreliable.  Then, 
conversely, the information dimension weights the data most uniformly, 
and so ``knows'' most about the fractal, without being overly 
sensitive to the possible occurrence of regions of very low 
probability.  The information dimension seems most informative.

\section{Fractal dimensions unbiased by finite size of data sets}
\label{ss3}
Cronin \& Roberts~\cite{Roberts95b} proposed a novel method to
eliminate biases, caused by finite sized data sets, in determining the
multi-fractal properties of a given data set.  Jelenik et
al.~\cite{Jelinek01, Jelinek04} used this method to explore the shape
of neuron cells.  The method compares characteristics of the inter-point
distances in the data set with those of artificially generated
multi-fractals.  By maximising the likelihood that the characteristics
are the same we model the multi-fractal nature of the data by the
parameters of the artificial multi-fractal.  By searching among
artificial multi-fractals with precisely the same number of sample
points as in the data, we anticipate that biases due to the finite
sample size will be statistically the same in the data and in the
artificial multi-fractals; hence predictions based upon the fitted
multi-fractal parameters should be unbiased by the finite sample size.

The method also appears to give a reliable indication of the error in
the estimates---a very desirable feature as also noted by Judd \&
Mees~\cite{Judd91}.  Most importantly for this paper, I generate finite
size data sets with specific parameters for the following specific
multiplicative multi-fractal process.  Given parameters
$\rho\in[0,0.5]$ and $\phi\in[0,0.5]$ a binary multiplicative
multi-fractal is generated by the recursive procedure of dividing each
interval into two halves, then assigning a fraction~$\phi$ of the
points in the interval to a random sub-interval of length~$\rho$ in the
left half, and the complementary fraction $\phi'=1-\phi$ to a random
sub-interval of length~$\rho$ in the right half.  Such a process has
generalised dimension
\begin{equation}
	D_q=\frac{\log\left(\phi^q+{\phi'}^q\right)}{(q-1)\log\rho}\,,
	\label{emdq}
\end{equation}
and a multi-fractal spectrum $f(\alpha)$~\cite[\S4]{Halsey86} given 
parameterically in terms of $0<\xi<1$ and $\xi'=1-\xi$ as
\begin{equation}
		f = \frac{\xi\log{\xi}+\xi'\log{\xi'}}
		        {\log{\rho}} 
		\,, \qquad
		\alpha = \frac{\xi\log{\phi}+\xi'\log{\phi'}}
		        {\log{\rho}} \,.
	\label{emfs}
\end{equation}
Here I chose $\rho=1/3$ and $\phi=1/4$ and sample the process with 
$N=100$; such a multi-fractal forms a finite data set whose 
parameters we need to estimate from the sample.

As explained in~\cite{Roberts95b}, we analyse such a sample by probing
it with \emph{exactly} the same multiplicative multi-fractal process,
and seek the best fit parameters.  Here the resulting estimate of the
original parameters is then in error \emph{only} due to the finite size
of the sample of the original multi-fractal process.  Because we fit
the data with a process which we know includes the one that generated
the data (a luxury rare in practise), there is no other
error.  Thus the spread in errors that we see is characteristic of only
the errors induced by a finite sized sample, nothing else.  In
particular,  observe that the deductions of the preceding section are
indeed appropriate.

\begin{figure}
\centerline{\includegraphics[width=0.95\textwidth]{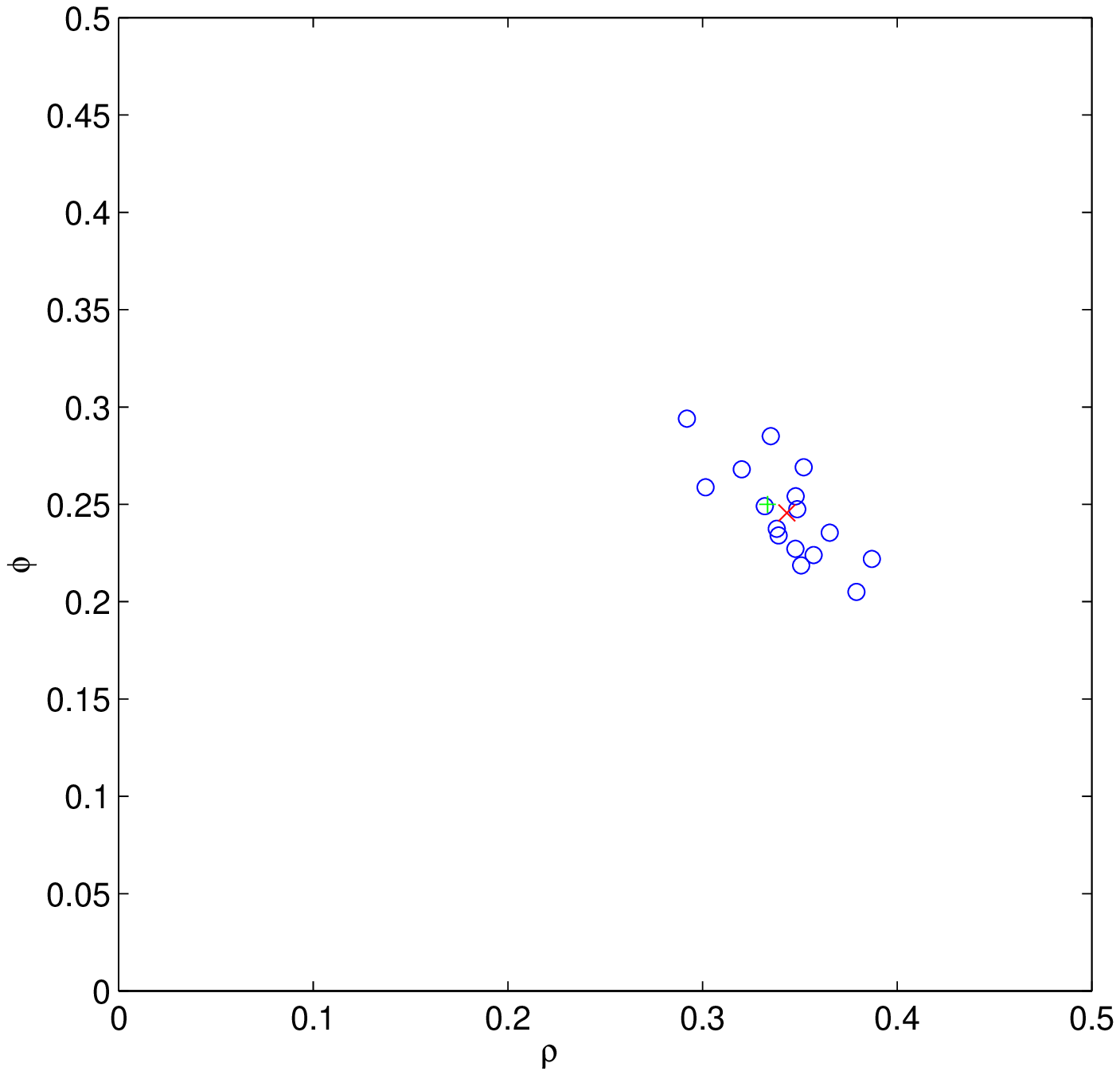}}
\caption{predicted multi-fractal parameters $(\rho,\phi)$, indicated 
by~$\circ$'s, from the maximum likelihood match to an ensemble of 16 
different realisations, each of $N=100$ data points, of a binary 
multiplicative multi-fractal with parameters $\rho=1/3$ and 
$\phi=1/4$, indicated by~$+$.  The mean location of the 
predictions is indicated by a~$\times$.}
\protect\label{n100means}
\end{figure}

I repeat the sampling of the multi-fractal followed by a maximum
likelihood estimate of the parameters 16~times.  Figure~\ref{n100means}
plots the estimates of the parameters.  Observe that the whole sampling
and estimation process appears unbiased in that the mean of the
predictions is reasonably close to the correct values of the
parameters.

\begin{figure}
\centerline{\includegraphics[width=0.95\textwidth]{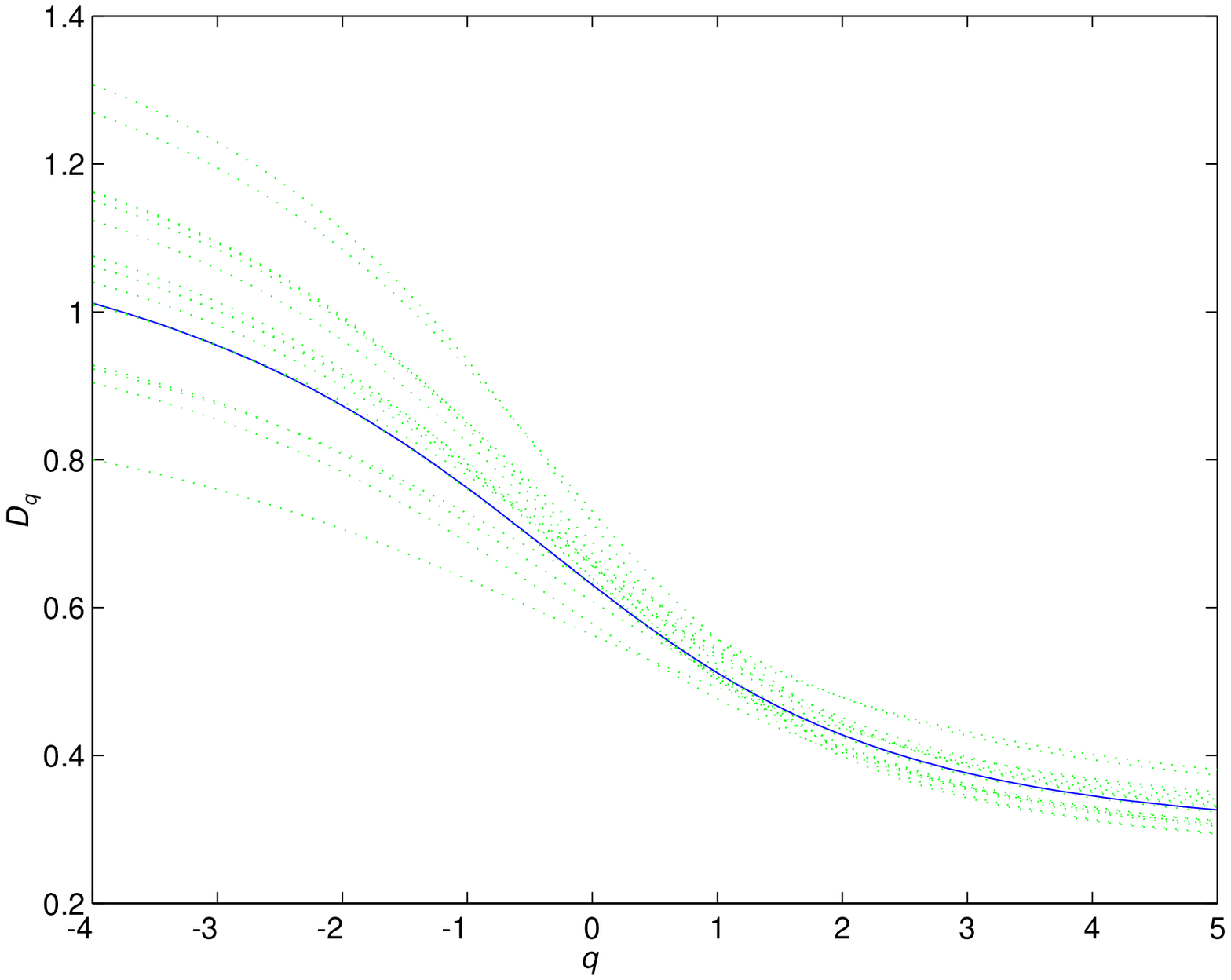}}
\caption{ensemble of multi-fractal generalised dimensions~$D_q$, 
dotted, for each of the predictions plotted in 
Figure~\protect\ref{n100means} made from samples of $N=100$ data 
points.  For comparison the generalised dimensions for the actual 
fractal is plotted as the solid line.  Observe the good estimation 
near the information dimension, but the large errors for negative 
order~$q$.}
\protect\label{n100dqs}
\end{figure} 

Ultimately, experimenters want to examine multi-fractal properties of 
the data.  Here these will be determined from the parameters 
$(\rho,\phi)$ of the best fit multi-fractal substituted into analytic 
expressions such as (\ref{emdq})~and~(\ref{emfs}).  For each of the
16~realisations and their best-fit estimates, I plot the corresponding 
predicted generalised dimensions~$D_q$ in Figure~\ref{n100dqs}.  (The 
corresponding graphs of the multi-fractal spectra~$f(\alpha)$ are 
plotted in Figure~7 of~\cite{Roberts95b} along with the true 
$f(\alpha)$~curve.)  Observe that the predicted dimensions for 
positive~$q$ (low~$\alpha$) are quite good for all realisations, 
especially near the information dimension,~$D_1$.  However, predicted 
dimensions for negative~$q$ (high~$\alpha$) are very poor; this is 
also the case for the Hausdorff dimension~$D_0$ (the maximum of the 
$f(\alpha)$~curve).  The negative~$q$ predictions are poor despite the 
fitting process ``knowing'' that there are no very low probability 
regions in this artificial process.  In general applications one 
cannot know this and I expect the negative~$q$ (large~$\alpha$) 
predictions to be significantly worse.  These numerical results 
convincingly support the arguments of the preceding section that we
should use the information dimension, not the Hausdorff.

\bibliographystyle{plain}
\bibliography{ajr,bib}

\end{document}